\documentclass[]{interact}

\usepackage{placeins}
\usepackage{epstopdf}
\usepackage{comment}
\usepackage[caption=false]{subfig}
\usepackage{xcolor}
\usepackage{braket}
\usepackage[numbers,sort&compress]{natbib}
\bibpunct[, ]{[}{]}{,}{n}{,}{,}
\renewcommand\bibfont{\fontsize{10}{12}\selectfont}
\makeatletter
\def\NAT@def@citea{\def\@citea{\NAT@separator}}
\makeatother

\theoremstyle{plain}

\theoremstyle{definition}

\theoremstyle{remark}

\begin{document}

\articletype{REVIEW ARTICLE}

\title{Quantum light microscopy}

\author{
\name{W.~P. Bowen\textsuperscript{a}\thanks{CONTACT W.~P. Bowen. Email: w.bowen@uq.edu.au}, Helen M. Chrzanowski\textsuperscript{b},  Dan Oron\textsuperscript{c}, Sven Ramelow\textsuperscript{b,d}, Dmitry Tabakaev\textsuperscript{e,f}, Alex Terrasson\textsuperscript{a} and Rob Thew\textsuperscript{e}}
\affil{\textsuperscript{a} Australian Research Council Centre of Excellence in Quantum Biotechnology, School of Mathematics and Physics, University of Queensland, Australia; 
\textsuperscript{b} Institut für Physik, Humboldt-Universität zu Berlin, Newtonstr. 15, 12489 Berlin, Germany;\textsuperscript{c} Dept. of Molecular Chemistry and Materials Science, Weizmann Institute of Science, Rehovot, Israel; \textsuperscript{d} IRIS Adlershof, Humboldt-Universität zu Berlin, Berlin, Germany \textsuperscript{e} Department of Applied Physics, University of Geneva, Geneva, Switzerland; \textsuperscript{f} Photonics Research Unit, Sensor systems division, Silicon Austria Labs, Villach, Austria}
}

\maketitle

\begin{abstract}

Much of our progress in understanding microscale biology has been powered by advances in microscopy. For instance, super-resolution microscopes allow the observation of biological structures at near-atomic-scale resolution, while multi-photon microscopes allow imaging deep into tissue. However, biological structures and dynamics still often remain out of reach of existing microscopes, with further advances in signal-to-noise, resolution and speed needed to access them. In many cases, the performance of microscopes is now limited by quantum effects -- such as noise due to the quantisation of light into photons or, for multi-photon microscopes, the low cross-section of multi-photon scattering. These limitations can be overcome by exploiting features of quantum mechanics such as entanglement. Quantum effects can also provide new ways to enhance the performance of microscopes, such as new super-resolution techniques and new techniques to image at difficult to reach wavelengths. This review provides an overview of these various ways in which quantum techniques can improve microscopy, including recent experimental progress. It  seeks to provide a realistic picture of what is possible, and what the constraints and opportunities are. 
\end{abstract}


\section{Introduction}

From their first use for scientific purposes in the seventeenth century, light microscopes dramatically changed our understanding of the microscopic world, enabling the discovery of bacteria, cells and parasites~\cite{locy1913early,porter1976antony,gest2004discovery}. In the centuries that followed, microscopes have been improved through many technological innovations, such as the development of compound microscopes in the nineteenth century and of phase contrast, fluorescence and confocal microscopes in the early-to-mid twentieth century~\cite{ball1966early,burch1942phase,lichtman2005fluorescence,nwaneshiudu2012introduction}. The discovery of lasers in the 1960s drove another leap in microscope technologies, with their higher light intensities enabling many new types of microscope, including super-resolution microscopes, light-sheet microscopes and multi-photon microscopes~\cite{klar2000fluorescence, rust2006sub,voie1993orthogonal,denk1990two,zipfel2003nonlinear}. 

Today, we are able to image biological specimens at near atomic scale~\cite{balzarotti2017nanometer,sigal2018visualizing,sahl2017fluorescence}, image the dynamics of cells in three dimensions and over long times~\cite{valm2017applying,lu2019lightsheet}, and image deep into highly scattering tissues~\cite{helmchen2005deep,huisken2004optical}. However, even with these impressive capabilities, the microscopic structures and dynamics that govern life remain in large part unobservable. While we can image with the far-sub-wavelength resolution relevant to sub-cellular processes, we generally cannot do this at the relevant timescales, rather averaging over the dynamics to retrieve a semi-static picture~\cite{schermelleh2019super}. Far-sub-wavelength resolution also remains challenging to achieve in uncontrolled environments, such as the interior of a living cell. Moreover, the  laser intensities required for precision microscopes are also, often, incompatible with biological specimens, so that weakly interacting and small structures are inaccessible without damaging the specimen~\cite{waldchen2015light,mauranyapin2017evanescent}.

Further advances in our understanding of microscale life demand microscopes that are more sensitive and have high resolution, while also operating with improved imaging speeds. Quantum microscopy provides a possible path to achieve these improvements. Quantum light microscopes exploit the laws of quantum physics to achieve performance that exceeds what is otherwise possible, or to enable entirely new forms of microscopy. For instance, quantum correlated photons can allow the noise floor of a microscope to be suppressed beneath the usual shot noise limit due to randomness in photon statistics~\cite{taylor2013biological,casacio2021quantum}. This can allow microscopy at lower intensities or higher speeds. Similarly, the fact that a fluorescent marker can only emit one photon at a time, combined with precision photon counting, provides information that enables new types of super-resolution microscopes~\cite{lelek2021single}. Photons of vastly different wavelength can also be quantum entangled, and this can be exploited to allow microscopy at hard to reach wavelengths such as the mid-infrared~\cite{kviatkovsky2020microscopy,kviatkovsky2022mid}.

With this review paper we aim to provide an accessible overview of quantum light microscopy, introducing the reader to some of the main approaches developed to date and what has been achieved, discussing what may be possible in future, giving a realistic assessment of challenges and limitations, and speculating on the potential areas of impact. The paper is not intended to be exhaustive, but rather focus on the key examples of sub-shot-noise microscopy, quantum multi-photon microscopy, microscopy with undetected photons, and quantum super-resolution microscopy. 
Several review papers have been written on the broader applications of quantum technologies in imaging and the life sciences, which we can recommend to the reader, along with references for key imaging techniques~\cite{casacio2021quantum,taylor2016quantum,giovannetti2004quantum,zhang2023quantum, li2022quantum,xu2022quantum,dorfman2016nonlinear, raymer21, lemos2014quantum,Santos2022,cui2013quantum, israel2017quantum,unternahrer2018super, tenne2019super,defienne2022pixel,mauranyapin2022quantum}.

\section{Microscopy beyond the shot noise limit}


\subsection{Precision limits of microscopy}


The clarity and speed of imaging is determined by the ratio of the signal strength to the noise level, or signal-to-noise. In microscopy, signals are generally weak since they arise from micro- or nanoscopic regions of a specimen. They are often enhanced using a contrast enhancing agent, such as a fluorescent dye~\cite{giepmans2006fluorescent}. Even so, it is common  to employ sophisticated strategies to reduce noise~\cite{ji2008advances,kumar2021speckle}. 

If all other noise sources are removed, the noise floor of a microscope is determined by the arrival statistics of photons on the imaging camera or detector. Ideally, for a low noise laser, these statistics are random, or Poissonian -- the arrival time of one photon provides no information about when subsequent photons will arrive. This results in optical {\it shot noise}. Due to its Poissonian nature, the standard deviation of shot noise is equal to the square-root of the number of photons that are detected within the measurement interval. By contrast, the measured signal typically scales linearly or, for a multi-photon microscope, super-linearly with the number of detected photons. A common strategy to increase signal-to-noise is, therefore, to increase the brightness of the microscope illumination. Alternatively, quantum correlations between photons can be used to reduce the optical noise floor beneath the shot noise limit.

Increasing illumination brightness has proved to be a highly effective strategy to enable precision microscopy~\cite{taylor2016quantum}. However, there are limits to its effectiveness. As a first challenge, as the shot noise-limited signal-to-noise increases, other noise sources such as acoustic noise, vibrational noise and laser frequency noise, tend to dominate. This can be seen from a simple example: determining the number of photons transmitted through a medium.  Here, we envisage a laser field passing through some absorbing medium, along the lines of absorption microscopy. We wish to  determine the mean transmittance of the medium as precisely as possible. Let us say that the transmittance of the medium is $T = \langle T \rangle + \delta T$, where $\delta T$ is a zero-mean fluctuating noise on the transmittance, for instance due to thermal fluctuations in the refractive index of the media that introduce light scattering.

We envisage that the transmitted field is perfectly detected with each photon converted into an electron. This generates the photocurrent $i=T n$. The light carries optical shot noise $\delta n$ due to its quantisation into photons with discrete energy, with $n = \langle n \rangle + \delta n$. Assuming that both the shot noise and transmittance noise are small in relative terms, we then have $i=(\langle T \rangle + \delta T)(\langle n \rangle + \delta n) \approx \langle T \rangle \langle n \rangle + \langle T \rangle \delta n + \langle n \rangle \delta T$.
The mean of this photocurrent, $\langle i \rangle =  \langle T \rangle \langle n \rangle$, provides a signal from which we can estimate the transmittance. The variance $V(i) = \langle i^2 \rangle - \langle i \rangle^2 = \langle T \rangle^2 V(\delta n) + \langle n \rangle^2 V(T)$ determines the accuracy of the estimate, and includes contributions from both shot noise and classical transmittance fluctuations.
Assuming the light is Poissonian, the shot noise component is equal to the mean number of detected  photons $\langle i \rangle$, i.e., $\langle T \rangle^2 V(\delta n) = \langle T \rangle \langle n \rangle$. 
The signal-to-noise of the measurement, defined conventionally as signal-squared divided by noise variance, is then
\begin{equation}
{\rm SNR } = \frac{\langle i \rangle^2}{V(i)} =  \frac{\langle T \rangle^2 \langle n \rangle}{\langle T \rangle + \langle n \rangle V(T)}. \label{basicSNR}
\end{equation}

From Eq.~(\ref{basicSNR}) an important point becomes clear -- while at low photon numbers ($\langle n \rangle \ll \langle T \rangle/V(T)$) the measurement is shot noise limited so that increasing the optical power increases the signal-to-noise linearly, at high photon numbers the signal-to-noise saturates at a constant value (at least in this case of a linear measurement) that is determined by classical noise ($SNR \approx \langle T \rangle^2/V(T)$ when $\langle n \rangle \gg \langle T \rangle/V(T)$). For this reason, a primary objective in precision microscopy has historically been to design configurations that naturally suppress classical noise sources. For instance, dark-field microscopes are configured so that the light that is used to illuminate the specimen does not directly impinge upon the imaging camera~\cite{gao2021dark,taylor2014subdiffraction, taylor2013enhanced}, suppressing background noise from illumination. Similarly, fluorescence and multi-photon microscopes collect light at a different frequency from the illumination field~\cite{lichtman2005fluorescence,konig2000multiphoton}; while differential interference contrast microscopes and polarisation microscopes function in essence by measuring phase shifts between two similar optical paths, and are insensitive to phase noise that is common to both paths~\cite{arnison2004linear}. 

A corollary to  Eq.~(\ref{basicSNR}) is that, if the goal is to maximise signal-to-noise and there are no constraints on the available laser power, classical noise sources will {\it always} in practice be the dominant source of noise in microscopy. In this case quantum correlations should not be expected to have a practical role to play in reducing the noise floor. Never-the-less, often, optical power levels are constrained, commonly by photo-intrusion on the specimen and sometimes by the power handling of the detectors used or limitations in available laser power~\cite{giovannetti2004quantum,ashkin2000history, carlton2010fast,pena2012optical}. When there is both a constraint on optical power, and the peak power that may be used in the microscope is shot noise limited, there is the prospect to use quantum correlations to provide a practical performance advantage.

While some common microscopy techniques, such as bright field microscopy, are generally not constrained by the illumination power as they only require moderate amount of light, many techniques are and have achieved shot-noise limited measurements, making them potential candidates for quantum enhancement. Three primary examples are fluorescence microscopy, interference-based microscopy, and nonlinear optical microscopy.

Fluorescence microscopy is the mainstay of microscopy in the field of biology thanks to its high contrast and intrinsic selectivity. As an example, the combination of fluorescence-probe technology and modern optical microscopes have allowed to investigate dynamics in living cells with high temporal and spatial resolution \cite{de2015frap,ettinger2014fluorescence,lippincott2003development}.
However, its illumination power is constrained by photobleaching and phototoxicity \cite{hoebe2007controlled,laissue2017assessing}, and various forms of fluorescence microscopy have reached the shot-noise limit~\cite{fujita2007high}. The constraint this introduces on the signal-to-noise ratio is particularly problematic for modalities that achieve high spatial resolution such as confocal microscopy~\cite{wilson1990confocal}, TPEF (two photons excited fluorescence) \cite{denk1990two,so2000two} and TIRF (total internal reflection microscopy) \cite{axelrod2001total}, and especially problematic for super-resolving microscopes such as PALM (photoactivated localization microscopy)~ \cite{lee2012counting}, STORM (stochastic optical reconstruction microscopy) \cite{rust2006sub}, STED (stimulated emission depletion microscopy)~\cite{hell1994breaking}, and FRET (fluorescence resonant energy transfer) \cite{selvin2000renaissance}. 

Interference-based microscopies leverage the phase information contained in a sample, for example due to changes in refractive index.
Two prominent techniques in this category are optical coherence microscopy  \cite{izatt1994optical} and interferometric scattering microscopy (iSCAT) \cite{ortega2012interferometric}. Both of these methods already operate at the shot-noise limit, making them suitable candidates for sub-shot-noise measurements. Another innovative approach, known as photonic force microscopy (PFM)~\cite{florin1997photonic}, utilizes the motion of a trapped particle in an optical tweezer to map the spatial structure of biological samples at length scales down to tens of nanometers. PFMs face power constraints and are limited by shot noise~\cite{rohrbach2004trapping,friese1999three}. The integration of quantum correlations have already been reported in PFM, resulting in a 14\% increase in spatial resolution compared to the shot-noise limit when mapping the spatial structure within cells \cite{taylor2014subdiffraction}.

Nonlinear optical microscopes harness nonlinear interactions between light and matter. They provide unique capabilities; for example, multi-photon processes such as second and third-order harmonic generation microscopy (SHG, THG)~\cite{campagnola2002three,barad1997nonlinear} allow imaging deep within biological tissues due to the nonlinear dependence of the interaction strength on intensity and the reduced absorption of near-infrared photons~\cite{konig2000multiphoton,sanderson2014fluorescence}. Coherent anti-Stokes Raman scattering (CARS)~\cite{zumbusch1999three}, stimulated Raman scattering (SRS)~\cite{nandakumar2009vibrational,freudiger2014stimulated} and stimulated Brillouin scattering (SBS)~\cite{ballmann2015stimulated} enable label-free fingerprinting of chemical components~\cite{krafft2012raman} and viscoelastic properties. Because the signal scales nonlinearly with the instantaneous light intensity, pulsed light is often used to improve the SNR. As a consequence, nonlinear microscopes are often damage constrained~\cite{li2020adaptive,waldchen2015light, mauranyapin2017evanescent, fu2006characterization,schermelleh2019super}. SHG, CARS and SRS have demonstrated shot-noise limited measurement \cite{dombeck2004optical,jurna2007shot,rock2013near}. This combined with the photon budget limit makes them natural candidates for quantum enhancement.

\subsection{Microscopy beneath the shot noise limit}

Once a microscope has reached the shot noise limit at intensities that are damaging to the specimens it observes, there are limited options to improve its signal-to-noise ratio (or imaging speed at fixed signal-to-noise ratio). Some improvements may be possible by engineering the microscope objectives or other optical components to increase numerical aperture or decrease losses. However, these sorts of improvements quickly reach the limit of diminishing returns. Alternatively, contrast enhancing agents can be used to increase the strength of the interaction between the specimen and light~\cite{wei2017super}. However, these are undesirable in many situations. e.g. removing the unlabelled nature of the microscope~\cite{wei2017super} or introducing toxicity~\cite{frangioni2003vivo}.

Beyond these approaches, the signal-to-noise can only be further improved using quantum correlations between photons to reduce the variance of the optical noise or to increase the signal strength as discussed in Section~3. As shown for the specific case of transmission measurement in Eq.~(\ref{basicSNR}), the signal-to-noise of a linear shot noise-limited measurement scales as the mean photon number $\langle n \rangle$ (setting $V(T)=0$ in Eq.~(\ref{basicSNR})). This is referred to as {\it shot-noise scaling}. Quantum correlated states of light such as squeezed~\cite{giovannetti2004quantum,pezze2008mach} and NOON states~\cite{higgins2009demonstrating} can achieve better scaling, and therefore better signal-to-noise ratios.


The best scaling that can be achieved for a linear interaction is an improvement of signal-to-noise with  $\langle n^{2} \rangle$~\cite{giovannetti2004quantum}. This is termed {\it Heisenberg scaling}. In principle, this can allow vast improvements. For instance, consider a measurement for which 1~mW of laser light is detected for one second. With a total of around  $10^{16}$ photons detected, Heisenberg scaling would allow a sixteen order-of-magnitude improvement in signal-to-noise. In reality, increasingly fragile quantum correlations are required to achieve Heisenberg scaling as the photon number increases~\cite{taylor2016quantum}, precluding such large enhancements. Even if the required quantum correlations can be generated, the inevitable optical losses in the measurement apparatus will degrade them. In the case of phase measurement, it has been shown that no matter what the input quantum correlations, the maximum factor of enhancement in signal-to-noise possible over a perfectly efficient shot-noise-limited measurement is given by $\eta/(1-\eta)$~\cite{demkowicz2012elusive}, where $\eta$ is the total efficiency of transmission through the system and detection. This result can be expected to hold for any linear quantum-enhanced measurement. Thus, practical linear quantum-enhanced measurements can be expected to show shot-noise limited scaling, but with a fixed factor of signal-to-noise improvement. Similar results apply for non-linear measurements, with the caveat that the nonlinearity changes how the signal (rather than the noise) scales with light intensity.


Several different classes of quantum light allow the shot-noise limit to be overcome. Particularly common are entangled photon pairs~\cite{hong1987measurement}, single photon states~\cite{lounis2005single}, NOON states~\cite{afek2010high}, and squeezed states~\cite{walls1983squeezed}. Entangled photon pairs, single photon states and noon states are commonly produced probabilistically via spontaneous parametric down-conversion~\cite{zhang2021spontaneous,bocquillon2009coherence}, although deterministic sources are also being developed~\cite{uppu2020scalable,muller2014demand}. A significant challenge for applications of these sources of quantum light to reduce noise is that they are generally very weak, with photon fluxes far below the usual fluxes used in precision microscopes~\cite{casacio2021quantum}. Squeezed states of light overcome this brightness challenge by combining quantum correlations between photons with a bright coherent field. They are typically produced in a three-wave mixing process such as optical parametric amplification.
The quantum correlations produced in this process can be used to redistribute noise between the amplitude and phase of the optical field (more precisely the amplitude and phase quadratures), allowing the noise in one to be reduced below the shot noise limit at the expense of increased noise in the other.

The ability to combine field brightness with quantum correlations explains why squeezed light is the quantum state of choice for gravitational wave detection \cite{aasi2013enhanced, mckenzie2004squeezing}. A second attractive property of squeezed states, is that the enhancement in noise floor that they provide naturally approaches the fundamental limit due to inefficiency. Consider a signal $s$ encoded on the amplitude quadrature $X$ of an optical field, so that $X = s + n$, where $n$ is the noise on the field. If this field was shot noise limited and detected with perfect efficiency, the signal-to-noise ratio would be $SNR = \langle X \rangle^2/\langle n^2 \rangle = s^2$, where we have normalised the shot noise to $\langle n^2 \rangle = 1$ for convenience. If, on the other hand, the field encountered some loss, then a fraction would be replaced with vacuum.
For a transmission efficiency of  $\eta$, $X \rightarrow \sqrt{\eta} X + \sqrt{1-\eta} v$, where $v$ is vacuum noise. The signal-to-noise ratio then becomes $SNR = \eta \langle  X \rangle^2/\langle (\sqrt{ \eta} n+ \sqrt {1-\eta} v)^2 \rangle = \eta s^2/(\eta V(n) + 1-\eta)$. If the field is strongly amplitude squeezed so that $V(n) \rightarrow 1$, this reduces to $SNR_{\rm sqz} =  \eta s^2/(1-\eta) = \eta/(1-\eta) \times SNR$, precisely the fundamental maximum enhancement possible for any state of light.

\subsection{Experimental demonstrations of microscopy beneath the shot noise limit}
Several types of quantum correlations have been used to enhance signal-to-noise in light microscopes. Spontaneous parametric down-conversion (SPDC) \cite{genovese2005research} has enabled the first demonstration of sub-shot noise imaging, making use of subtraction of the quantum correlated noise pattern to image absorbing samples \cite{brida2010experimental}. Following this work, SPDC has been utilized for imaging weakly absorbing samples using phase-contrast polarization microscopy \cite{ono2013entanglement,morris2015imaging,li2018enhanced,zhang2023quantum, ortolano2023quantum}, wide-field imaging \cite{samantaray2017realization}, transmittance optical microscopy \cite{sabines2019twin}, and two-photon absorption fluorescence microscopy \cite{Varnavski2020}.

NOON states, which are quantum superpositions of photons in different polarization modes \cite{afek2010high}, have been successfully integrated into phase-contrast microscopes ~\cite{ono2013entanglement,israel2014supersensitive}.
Squeezed light has been applied in a range of imaging techniques: probe scanning microscopy \cite{taylor2014subdiffraction}, stimulated emission microscopy \cite{triginer2020quantum}, vibrational microscopy such as Raman microscopy \cite{de2020quantum, casacio2021quantum,xu2022stimulated} and Brillouin microscopy \cite{li2022quantum}.

\begin{figure}[h]
	\centering
	\includegraphics[width=0.80\columnwidth]{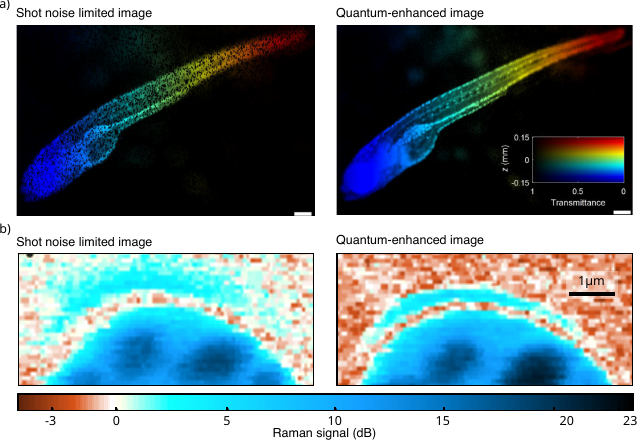}
	\caption{a) Comparison of quantum-enhanced (right) and classical (left) ICE transmittance images of a zebrafish in presence of stray light. The technique reports 25 times less sensitivity to stray light and allows improved image quality. Scale bars: 200$\mu m$. Adapted from \cite{zhang2023quantum}, with permission from the authors. b) Comparison of classical (left) and quantum-enhanced (right) SRS images of a yeast cell. The reduced noise floor appears in green in the quantum-enhanced image and allows sharper images by providing a 34\% increase in SNR. Adapted from \cite{casacio2021quantum} with permission from the authors.}
	\label{sub_shot_noise_images} 
\end{figure}

Among the demonstrations mentioned above, we specifically focus on experiments that applied quantum light to imaging biological samples.
Sub-shot noise microscopy of biological specimens was reported in 2013~\cite{taylor2013biological,taylor2014subdiffraction}. This work uses squeezed light to reduce the noise in particle tracking within cells, enabling quantum-enhanced estimation of the local viscoelasticity of the cell~\cite{taylor2013biological}. Employed in PFM it allowed the acquisition of one-dimensional profiles of spatial structures within a cell at scales down to 10 nm~\cite{taylor2014subdiffraction}, with the squeezed light yielding a 14\% enhancement in spatial resolution over coherent light. 
In 2015, quantum-enhanced imaging of a weakly absorbing sample -- a wasp wing, in this case -- was reported~\cite{morris2015imaging}. In this work, entangled photon pairs generated through SPDC were split into two arms. The photons detected on one arm were used to predict the arrival time of photons in the other arm, in a coincidence counting strategy. This approach, combined with time-gating of a CCD camera, significantly reduced background counts, enhancing the SNR. High-quality images were produced with fewer than one photon detected per pixel, a regime where the noise would dominate under classical illumination.
In another study \cite{zhang2023quantum}, sub-shot noise images were produced utilizing spatially and polarization entangled photon pairs combined with coincidence counting method. The technique allows increase SNR compared to classical illumination thanks to sub-shot noise operation and stray light suppression.
The technique was applied to image zebrafish, as shown in Fig.~\ref{sub_shot_noise_images}, panel a). The quantum image on the right exhibits significantly greater resilience to stray light compared to the classical image on the left. 

In the previous section, we discussed the primary motivation behind introducing quantum correlations to light microscopy: enhancing the SNR and speed in microscopes already constrained by shot noise and photodamage. The first demonstration of sub-shot noise light microscopy at the photodamage limit was achieved in Ref.~\cite{casacio2021quantum}. This work integrated a bright squeezed-light source into a stimulated Raman microscope, a type of microscope utilizing stimulated Raman scattering (SRS) to probe the vibrational modes of molecules. Such microscopes provide high sensitivity, specificity, and label-free biological measurements \cite{cheng2015vibrational,wei2017super}. The use of squeezed light allowed a 34\% increase in SNR in imaging of living yeast cells. Notably, even a small increase in laser illumination intensity was shown to introduce irreversible damage to the cells. This represents an absolute quantum advantage: using the same mode shapes and apparatus, the quantum SNR cannot be reproduced classically. It  allowed cell features to be resolved that were not visible using coherent illumination. This can be seen in Fig.~\ref{sub_shot_noise_images}(b), which  compares  cell images acquired with squeezed light (right) and the equivalent shot-noise limited  light (left). 

Following the demonstration of quantum enhanced microscopy at the photodamage limit in Ref.~\cite{casacio2021quantum}, recent works have further advanced the state-of-the-art of quantum-enhanced vibrational microscopy. Refs.~\cite{xu2022stimulated} and \cite{xu2022quantum} use balanced detection to demonstrate high detected powers (respectively 11.3mW and 34mW) in a quantum enhanced SRS microscope. Ref.~\cite{xu2023dual} also performs quantum enhanced SRS imaging, and is able to simultaneously detect the Raman signal along two polarizations, demonstrating quantum-enhanced multidimensional SRS measurement. Ref.~\cite{li2022quantum} reports sub-shot noise measurement at high power in stimulated Brillouin scattering, a technique used to image viscoelastical properties of a sample with applications for biologists \cite{vaughan1980brillouin,bausch2006bottom}. 3.4dB of enhancement and quantum enhanced images of water as the signal medium is demonstrated, with the next step to apply the technique to biological samples. 

We note that the level of SNR improvement in each of these experiments remains modest due to low optical efficiency and input squeezing levels. Improving this enhancement is an important challenge. A second open challenge is to combine quantum enhancement with other capabilities of vibrational imaging such as multispectral imaging~\cite{wei2017super}, video rate imaging~\cite{saar2010video}, and surface-enhanced Raman imaging~\cite{langer2019present}.

\vspace{2cm}

\FloatBarrier

\FloatBarrier
%
%
%
%
%
%
%

\section{Quantum multi-photon microscopy}

\subsection{Classical two-photon absorption}

Two-photon microscopy has been built on the seminal work of M. Goeppert Mayer~\cite{goppert2009elementary} describing two-photon absorption (TPA) by atoms. Following the invention of the laser, this process has been tested and refined, in multiple domains of research such as medicine, materials and life sciences, eventually finding its way to industrial applications. TPA is a nonlinear process whereby two photons are required to bridge the energy gap between ground and excited states - interestingly, this can be achieved without the presence of an intermediate state through what is often referred to as a ``virtual state" transition. Typically, these experiments require high-power lasers with short pulses to overcome low absorption efficiencies (low two-photon cross-sections). However, in the late 90s people started to wonder what would happen now if these two photons are entangled - could there be a quantum advantage? The theory pointed to potentially orders of magnitude lower photon fluxes being required if the photons were entangled. Again, it took some time, but experiments are starting to catch up with theory for entangled two-photon absorption (ETPA) and in the following we will look at some of the advantages, disadvantages, and challenges of this approach for non-linear microscopy.

\begin{figure}[h]
	\centering
	\includegraphics[width=0.5\columnwidth]{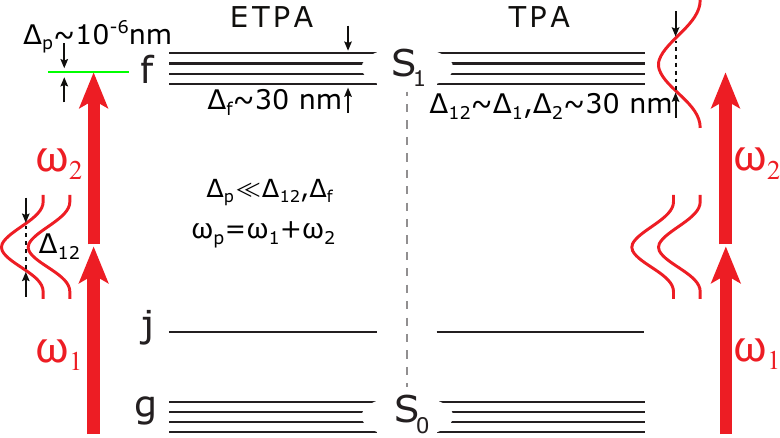}
	\caption{Simplified representation of the energy levels of an absorbing specimen showing and the various bandwidths of relevance. $\Delta_{1}, \Delta_{2}$ -- bandwidth of the laser emission in case of TPA; $\Delta_\text{f}$ -- bandwidth of the final excited state transition; $\Delta_\text{p}$ -- bandwidth of a pump laser with a circular frequency $\omega_p$, producing pairs with photons $\omega_1$ and $\omega_2$ (see next subsection); $\Delta_{12}$ -- bandwidth of a pair. Figure adapted from~\cite{tabakaev2021energy}.} 
	\label{fig:Energy_diagram} 
\end{figure}

The complete derivation of the multi-photon absorption formalism can be found at~\cite{bebb1966multiphoton, mollow1968two}, although here we limit ourselves to the so-called probabilistic model, which provides a mechanistic, yet illustrative,  picture; we consider interaction of three particles - one absorber, which can be either a molecule or atom, and two photons. The energy conservation law and Bohr model require that these two photons can be absorbed if and only if the sum of their energy fits the energy gap between ground and excited states of the absorber;  Fig.~\ref{fig:Energy_diagram} represents a simplified picture of the various ground and excited state energy level manifolds, virtual levels and their associated bandwidths, along with the bandwidths for laser and entangled pair sources. If the absorber can go back to the ground state, emitting a photon, then in the case of fluorescence or second-harmonic generation microscopy, this emitted photon is well separated spectrally from the excitation photons. This provides the first advantage of TPA microscopy -- simplified spectral filtering of the signal and higher signal-to-noise ratio (SNR), in comparison to single-photon absorption (SPA) techniques. 

Another advantage for ETPA lies in the fact that, while both classical and quantum schemes probe the intermediate levels with $fs$ bandwidth photons, ETPA provides a very sharp spectral resolution as the energy correlations ensure the pairs' bandwidth corresponds to the original pump laser and not the convolution of two photons from a pulsed system. While this may provide some improvement in efficiency for absorption, it is not typically orders of magnitude. Furthermore, as the Rayleigh scattering cross-section is inversely proportional to the wavelength of light, it is clear that less-energetic photons are scattered less, which means that when focused, they preserve the minimal size of the focal spot deeper into a sample, increasing the imaging depth.

In the following we introduce classical TPA, its extension to ETPA and the connection to an ETPA cross-section. We then discuss temporal and spatial characteristics that are relevant, e.g., for microscopy. We start by considering the two-photon absorption rate of a photon flux of $\phi$ [cm$^{-2}$s$^{-1}$]~\cite{fei1997entanglement, dayan2007theory, schlawin2017entangled}
\begin{equation}
	\label{eq:R_2}
	R_{2} = \delta \phi^2,
\end{equation}

The square dependence of the absorption rate on the excitation flux comes from the fact that arrival time of one photon is not linked to the arrival time of the second photon, so the probability to have two photons and one absorber somewhere within a space-time volume of $\delta$ is a product of two independent event probabilities of having one photon and another one. $\delta$ [cm$^{4}$s] is a two-photon absorption cross-section, and can be seen as a material-specific efficiency of the process. Its typical values are on the order of 10$^{-50}$ cm$^{4}$\,s~\cite{sperber1986s}. Such a small value requires one to increase $\phi$ by using ps or fs laser pulses, concentrating the photons temporally, and focusing the pulses down to $\mu$m areas, resulting in an observable signal only in the focus, the point of highest intensity. This latter point also opens up possibilities for 3D imaging.

However, using fs-lasers and strong focusing conditions to compensate for process inefficiency results in exposure of a sample to megawatts per cm$^2$~\cite{van2011action}, which creates the risk of photo-damage, especially for sensitive samples in medicine and life sciences.

\subsection{Entangled two-photon absorption}

In the late 90s the first theoretical studies examined what would happen if the two photons involved in the TPA process were entangled~\cite{fei1997entanglement}. After some time, this question was revisited and efforts continue to better understand and describe this interaction~\cite{dayan2007theory, schlawin2017entangled, dorfman2016nonlinear}. Entanglement provides the intuitive advantage that if one photon arrived on a given molecule at a certain time, the second one should also arrive. Indeed, if we imagine that the absorber is excited by pairs of photons with flux $\phi_e$, then the pair absorption rate becomes linear as a function of $\phi$
\begin{equation}
	\label{eq:R_e}
	R_\text{e}=\sigma_\text{e}\phi,
\end{equation}

where $\sigma_{e}$ is the entangled two-photon absorption (ETPA) cross-section, which plays the role of pair absorption efficiency and has units of a regular single-photon absorption cross-section [cm$^{2}$]. It can be linked to the classical TPA cross-section in the following way~\cite{raymer2021tutorial, landes2020experimental, raymer21}:
\begin{equation}
	\begin {split}
	\label{eq:EtpaCrossSection}
	\sigma_{e} = f\frac{\delta}{2A_eT_c}.
\end{split}
\end{equation}

We immediately see the dependence on the $T_c$, the so-called entanglement time, or more generally, the coherence time, as well as the entanglement area $A_e$, which is defined as an uncertainty of the pair emission position inside the crystal~\cite{jost1998spatial}. The factor $f$ is a quantum-enhancement factor, which has recently been introduced to take into account the degree of correlations in entangled pairs as a function of their source characteristics, but it can also be used to capture other, possibly unknown, enhancement factors. This $f$ can be quite large, e.g., it has an the order of magnitude of around $10^8 - 10^9$, with respect to \cite{tabakaev2022spatial}.  Importantly, we can see that the efficiency of ETPA becomes not just material dependent, but also depends on the photon pair source properties.
 
There are several techniques to experimentally generate entangled photon pairs and different degrees of freedom that can be entangled, but the most popular technique is through spontaneous parametric down-conversion (SPDC) - a nonlinear process, within which a pump laser photon is destroyed and a pair of photons (typically called signal and idler) are created, and where the sum of their energies is equal to that of the pump photon. This process also guarantees that photons of the same pair are generated simultaneously and the process is said to produce energy-time entangled photon pairs, i.e., they are correlated in their generation time and anti-correlated in energy. We further limit ourselves to energy-time entangled pairs only, as other degrees of freedom have not yet demonstrated any apparent influence on two-photon absorption efficiencies~\cite{tabakaev2021energy}.

\subsection{ETPA cross-section}

\begin{figure}
	\centering
	\includegraphics[width=0.4\columnwidth]{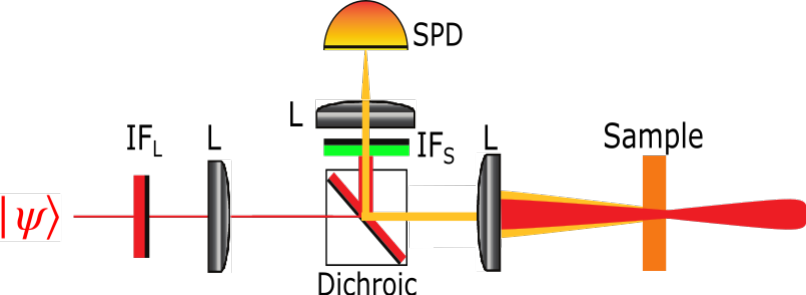}\hspace{5mm}
    \includegraphics[width=0.4\columnwidth]{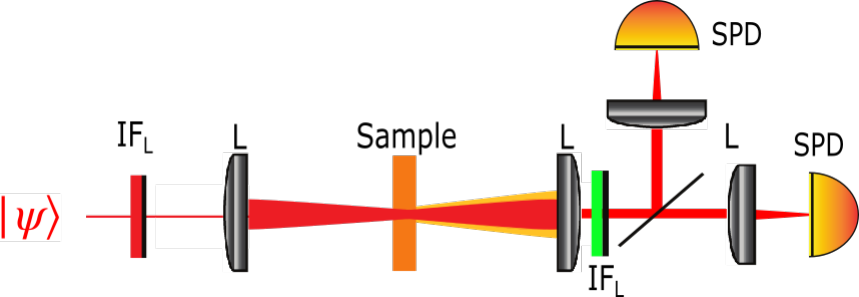}
	\caption{(Left) Typical ETPA fluoresence measurement scheme injecting entangled photons and detecting the back-reflected fluorescence: L -- lens; IF$_\text{L}$ -- low pass filter(s); IF$_\text{s}$ -- short-pass filter(s); SPD -- single-photon detector. (right) Typical ETPA transmission scheme using SPD to detect singles and coincidence counts to monitor loss due to ETPA. The ''sample" is typically a cuvette holding the molecular solution under test.} 
	\label{fig:scheme} 
\end{figure}

If we now focus on the cross-section, and consider Eq.~\ref{eq:R_e}, we can see that $\sigma_{e}$ defines a linear relationship between the number of detected counts and the input photon-pair flux. To measure this experimentally, the linearity measurements were performed by several groups, where they varied the input pair flux and detected either the ETPA-induced fluorescence~\cite{lee2006entangled, dayan2004two, dayan2005nonlinear, harpham2009thiophene, upton2013optically, varnavski2017entangled,villabona2020measurements, tabakaev2021energy} or the change in the number of  pairs, in transmission, after they pass through the sample~\cite{villabona2017entangled, villabona2020measurements}. Some typical values are presented in Table~\ref{tab_cross}. The values for Rh6G dye and Zinc tetraphenylporphine (Zn TPP) organic semiconductor were measured using the fluorescence detection protocol, while, for RhB dye, by photon-pair detection in transmission. The spread in cross-section values corresponds partially to different concentrations, however, there is still quite some variation in values that is currently not understood. Typical ETPA linearity measurement schemes are shown in Fig.~\ref{fig:scheme}. 

In these cases where values for ETPA cross-sections have been reported, there remain concerns about their validity, in particular, as to whether competing signals can be ruled out. In the case of Zn TPP cross-sections almost as large as for single-photon absorption have been reported, and similarly, some other studies~\cite{lee2006entangled, harpham2009thiophene, varnavski2017entangled} have demonstrated that ETPA can be as efficient as SPA for different compounds. One significant challenge is to rule out spurious signals that could be due to other processes such as pump leakage, which would produce SPA-induced fluorescence, which was a concern for some~\cite{lee2006entangled, lee2007quantum, harpham2009thiophene, upton2013optically, varnavski2017entangled, guzman2010organic, guzman2010spatial}. More recently, the  possibility of hot-band absorption~\cite{mikhaylov2022hot} has also been raised as a potential source or error in fluorescence experiments, such as~\cite{tabakaev2021energy}. To rule this out completely, temperature dependent measurements would need to be made. 

\begin{table} [h]
	\tbl{Some measured values of $\sigma_{e}$.}
	{\begin{tabular}{lccc} \toprule		
			Compound & $\sigma_{e}$, [cm$^{-2}$]  & Method & Ref. \\ \midrule
			Rh6G & $10^{-21}$ & Fluorescence & \cite{tabakaev2021energy} \\
			RhB & $10^{-18}-10^{-21}$  & Transmission  & \cite{villabona2017entangled}  \\
			Zinc tetraphenylporphine & $10^{-18}$  & Fluorescence  & \cite{upton2013optically} \\ \bottomrule
	\end{tabular}}
	\label{tab_cross}
\end{table}

It should be noted that in some experimental~\cite{corona2021experimental, parzuchowski2021setting, landes2020experimental, mikhaylov2020comprehensive} and theoretical~\cite{landes2021quantifying, raymer21} studies, bounds were put on the ability to observe signals produced by ETPA, both for fluorescence-induced ETPA or transmission experiments where the photon-pair losses are related to the process. Even for experiments that have claimed to observe ETPA signals, there are very few conclusive results and much work is still required to consistently reproduce these experiments. As such, there are widespread efforts across the community to better understand the theoretical models and the details of the experimental schemes and remove the current uncertainty that this factor $f$ in Eq.~\ref{eq:EtpaCrossSection} captures.

The main challenge for experimental ETPA demonstrations has been to distinguish ETPA signal from the much more efficient SPA, or direct detection of the laser or even of photon-pairs by a detector - the detector itself can undergo a two-photon absorption with non-negligible efficiency. An intuitive and elegant way to differentiate the two-photon process from the single-photon process was demonstrated in~\cite{dayan2005nonlinear}; if linear loss is introduced into the beam of photon-pairs, than the ETPA signal should scale quadratically as a function of the number of photon-pairs, because loss of any photon is equal to loss of the whole pair, in terms of the absorption process. In contrast, introducing the loss on the laser beam generating the photon-pairs, changes the ETPA linearly as only the number of pairs created is changed as an integer number and not by "half"-pairs. This test is generally considered to be the most robust in the community and has been demonstrated for a Rh6G molecular systems~\cite{tabakaev2022spatial} -- the value of $\sigma_{e}$ is given in  Table~\ref{tab_cross}. Unfortunately, as mentioned, this result has yet to be reproduced.

\subsection{Spatial and temporal properties of ETPA}

So far we have discussed the efficiency aspect of ETPA. Now we consider its spatial properties to better understand how this compares to classical single- and two-photon absorption. To do this, we must first go beyond the single molecule case and consider an effective excitation volume $A\,l$ and molecular concentration $C$. We also assume that the excited area $A$ is equal to the entanglement area $A_e$, which is an assumption that appears valid for the majority of experimental conditions~\cite{tabakaev2022spatial}. By carefully analyzing Eq.~\ref{eq:R_e} and taking into account the relation between $\sigma_{e}$ to $\delta$ we can see that ETPA rates should scale linearly with the entanglement time and the excitation area $A$, 
\begin{equation}
	\label{eq:R_espace}
	R_\text{ETPA} = C\,A\,l\,\frac{\delta}{A T} \frac{R_{pair}}{A}\,.
\end{equation}
This was recently experimentally examined by performing a Z-scan sectioning of thin layers of Rh6G showing that ETPA-induced fluorescence scales like TPA, thus potentially providing similar spatial resolution to TPA techniques~\cite{tabakaev2022spatial}.

To demonstrate the temporal properties of ETPA, its rate was measured as a function of temporal delay between the signal and idler fields both in CW and ns-pulsed pump regimes \cite{dayan2004two, tabakaev2022spatial}. Interestingly, this dependence was the same as if the ETPA transition was driven by classical pulses of the same duration as of signal and idler, demonstrating a sharp peak-like dependence with FWHM corresponding to the coherence -- or entanglement -- time of the photons. What differentiates ETPA from TPA is that this fs-scale resolution allows for simultaneous high spectral resolution, which is defined not by the wavelengths and bandwidths of signal and idler, but by the wavelength and bandwidth of the pump laser, which  can be tunable and, especially in the case of CW pumping, on the order of few MHz.

\subsection{Conclusion}

The ETPA process certainly provides a promising tool for low phototoxicity imaging due to high absorption efficiency of entangled photons in comparison to classical two-photon absorption, as well as appearing to have better spatial and temporal resolution than SPA. However, this approach currently suffers from low excitation rates compared to the efficiency of absorption resulting in the low signal - in any paper with positive results mentioned in this review, the detected signal is not more than 10 counts/s. The low signal to noise currently achieved, along with a number of detailed studies showing the limitations, even the impossibility of seeing meaningful signals, is driving work on better understanding ETPA more from a fundamental perspective but with an eye to what can be achieved in applied technique such as ETPA microscopy. First results demonstrating ETPA microscopy have been realised~\cite{varnavski2020two}, although again, this has not been reproduced and the how the results were achieved with such low photon pair fluxes requires further understanding.

The main challenges for ETPA feasibility as an applied technique, and more generally for its experimental studies, is the very low SNR, which is fundamentally related to the rate of entangled pairs, and there is a lack of connection between the observed results and the modern quantum optics formalism~\cite{raymer21}. Increasing the SNR can be achieved by increasing the bandwidth and consequently the rate of entangled pairs~\cite{szoke2021designing}, improving the collection efficiency for fluorescence-detection schemes and decreasing the detection noise level by using superconducting nanowire single-photon detectors~\cite{parzuchowski2021setting}  and imaging devices, could also help. There are ideas to consider chirped pulses and pairs, and generally, better understand the role of the source statistics in such experiments.
Perhaps the most important need is to establish a better theoretical understanding of ETPA process. This will require significant interdisciplinary research involving the light-matter interaction description not only from quantum optics but also from physical chemistry point of view, and many groups are already working in this direction. 

It is a rare thing in quantum optics to have such a disconnect between theory and experiment, and to have so many unknowns, but we are now beyond our relatively simple atomic systems and working with more complex molecules. Bridging the gap will require a focused effort, but perhaps in doing so, we will find a clear path to exploiting ETPA for microscopy and other applications that has the much promised quantum advantage that started this all several decades ago.

%
%
%
%
%
%
%
%

\section{Imaging and microscopy with undetected photons} 

\subsection{Induced coherence without stimulated emission \& non-linear inteferometry}

\begin{figure}[h]
	\centering
	\includegraphics[width=\columnwidth]{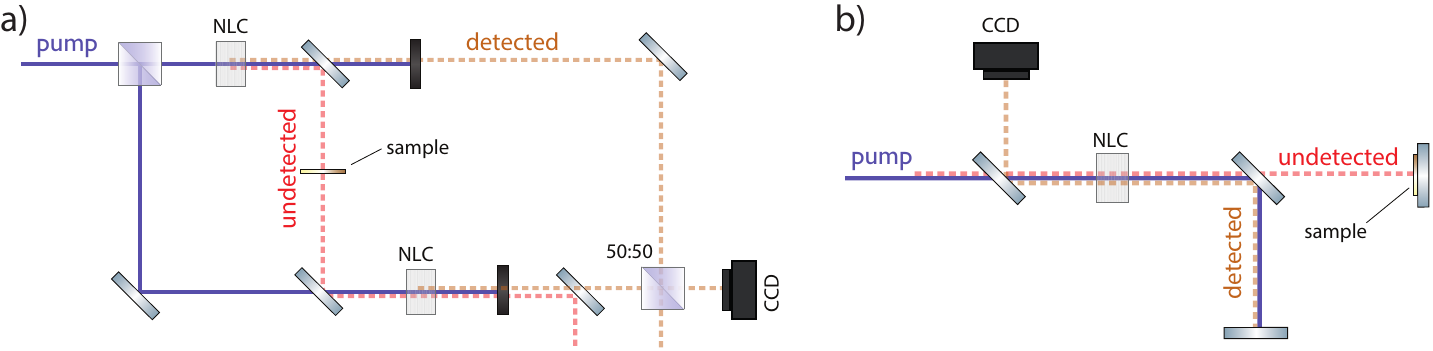}
	\caption{{\bf The two distinct interferometer schemes for quantum imaging with undetected photons:} {\bf a).} The imaging analog of the original Wang-Zho-Mandel interferometer~\cite{Wang91}. The undetected field emerging from the first non-linear crystal (NLC) is aligned into the second crystal. The detected fields emerging from both crystals are interfered on a 50:50 beam-splitter before measurement. {\bf b).} Alternative scheme based on a SU(1,1) non-linear interferometer~\cite{Chekhova:16}, all fields emerging from the first crystal are aligned into the second crystal.} 
	\label{fig:inteferometerschemes} 
\end{figure}

Microscopy (and imaging) with undetected photons is a relatively new technique which exploits quantum effects to decouple the sensing and detection wavelengths. The primary basis for microscopy, imaging and all other sensing modalities that exploit undetected photons is the phenomenon of `induced coherence without stimulated emission' \cite{Wang91}, which can also be framed in the language of nonlinear interferometers \cite{Chekhova:16}. The work of \cite{Wang91} was one amongst a series of seminal papers, in which Mandel and his co-authors probed the fundamental repercussions of the quantisation of light. The original experiment, sketched in Fig.~\ref{fig:inteferometerschemes}~a), considered two spontaneous parametric down conversion (SPDC) crystals coherently pumped in series, with the signal photon emerging from the first crystal aligned into the second crystal to ensure (ideally) perfect mode overlap with the photon emitted from the second crystal. The two resulting idler fields are overlapped together on a beamsplitter to ensure their spatial interference, and the intensity at its output ports is subsequently measured. A click at the photon detector would herald the generation of an idler photon, but without revealing whether it was produced in the first or the second crystal. This absence of `which source' information produces an interference effect in the measured idler intensity that depends on the relative phase accumulated between the pump and the signal field between the first and second crystal. Critically, this interference remains, with or without measurement of the signal light, which can be happily discarded after the second crystal. This effect of `induced coherence' between the first and second SPDC process degrades when loss is introduced in the path taken by the signal field between the first and second crystal, essentially revealing information about whether the bi-photon was born in the first or second crystal. 

It is important to note that a very similar phenomenon occurs if both fields (signal and idler) are aligned into the second crystal (Fig.~\ref{fig:inteferometerschemes}~b)). This particular variant of nonlinear interferometery~\cite{Chekhova:16}, where the beam-splitters coupling the modes of a typical linear interferometer are replaced by driven non-linear processes, was first introduced by Yurke {\it et al.} who denoted it SU(1,1) in reference to its symmetry group~\cite{Yurke86}. In wider quantum optics, the SU(1,1) interferometer is a well-studied device, notably in the topic of phase-sensing, where it potentially achieves Heisenberg scaling~\cite{Yurke86}.  

The field of quantum imaging with undetected photons (QIUP), in its modern incarnation, started with the experiment of Lemos {\it et al.} \cite{lemos2014quantum}. Twenty years on from the work of \cite{Wang91}, the authors revisited the aforedescribed experiment, but their innovation was to make use of the plentiful spatial entanglement present in SPDC to implement an imaging task. While \cite{lemos2014quantum} itself is largely presented as a work of quantum foundations, it brought two important insights: firstly, such interferometric techniques could leverage the abundant spatial entanglement present in SPDC to realise wide-field imaging. These spatial correlations present in SPDC had already seen extensive investigation in the context of intensity correlation imaging, for example, quantum ghost imaging \cite{aspden2013epr} and spatial analogues of Einstein-Poldosky-Rosen demonstrations \cite{howell2004realization}. Secondly, by using widely non-degenerate down-conversion processes, QIUP offered the possibility to move the illumination wavelength into regions either unserved or underserved by existing camera technology. With an appropriately designed SPDC source, widefield imaging in regions such as the mid- or far-IR, would be feasible with off-the-shelf CCD and CMOS imaging technologies. 

\subsection{Widefield imaging and microscopy implementations}

There are predominantly two approaches to QIUP. The first (Fig.~\ref{fig:inteferometerschemes}~a)) is the imaging analogue of the original Mandel experiment \cite{Wang91}, where only the `undetected' field enters the second crystal and the two `detected' fields emerging from the two crystals are subsequently interfered on a 50:50 beamsplitter and imaged \cite{lemos2014quantum}. In the second, and more widely favoured approach \cite{Paterova:2020,Paterova:2020ei,kviatkovsky2020microscopy,Gilaberte:21}, the two crystals form a traditional non-linear interferometer, with both fields produced in the first crystal passing through the second crystal (Fig.~\ref{fig:inteferometerschemes}~b)). Here, we give the simple theory of the latter, but the results are qualitatively equivalent for both approaches. 

The most commonplace approach is the implementation of a non-linear Michelson-type interferometer as detailed in Fig.~\ref{fig:inteferometerschemes}~b, owing to its compact folded geometry which uses a single crystal, and simplifies alignment and mode-matching. Upon the first pass of the non-linear crystal, the SPDC process produces two fields we label the {\it detected} and {\it undetected}, denoting their future roles. The two fields are subsequently split spatially (with the pump typically co-propogating with the detected light) and the undetected light illuminates the sample. All fields are reflected and aligned back into the crystal, after which their joint quantum state bears both the spatially dependent phase and absorption information of the sample. With the modes ideally aligned and assuming a balanced inteferometer, the `detected' light emerging from the crystal is detected via a spatial-resolved detection, with a measured intensity 
\begin{align}
I(r) \propto 1 + \sqrt{\eta_u(r)} \cos{(\Delta \phi (r) + \phi_0})
\end{align}
where $r$ labels a spatial coordinate, $\eta_u (r)$ is the transmission of the undetected field, and $\Delta \phi (r)$ is the relative phase difference accumulated between all three fields inside the interferometer. The visibility measured per spatial coordinate is directly proportional to the square-root of the loss experienced by the undetected field, allowing for the measurement of spatially dependent absorption at the undetected wavelength. The spatially dependent phase-shift imparted by the sample is also resolvable e.g. via scanning one of the interferometer end mirrors. 
\begin{figure}[h]
	\centering
	\includegraphics[width=\columnwidth]{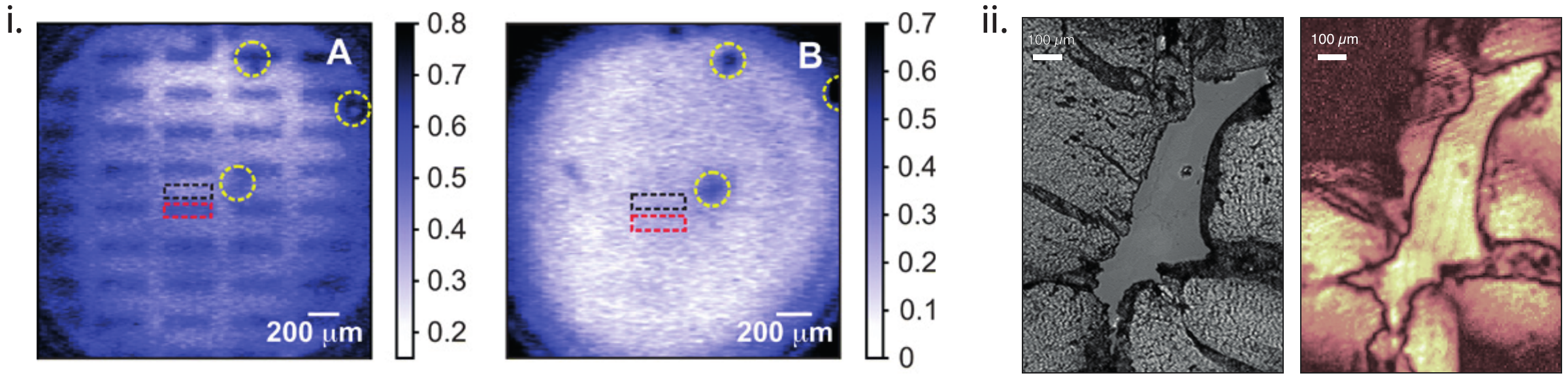}
	\caption{{\bf Microscopy with Undetected Photons} {\bf i).} Hyper-spectral microscopic images of a photosensitive polymer in the mid-IR using momentum (far field) correlations. Absorption image of the sample at undetected (detected) wavelength of (A) 2.87 $\mu$m (653.2 nm), (B) 3.18 $\mu$m (638.8 nm). {\bf ii.)} Classical white light (left) and undetected photon microscopy (right) of a mouse-heart cross section. The undetected microscopy was realised with position (image plane) correlations with an illumination (detection) wavelength of 3.8 $\mu$m (800 nm). Images adapted with permission of the authors from \cite{Paterova:2020} and \cite{kviatkovsky2022mid}.} 
	\label{fig:microscopyimages} 
\end{figure}
\subsubsection*{Far Field Imaging}
Almost all experimental implementations of QIUP have been realised with the sample and camera placed in the Fourier plane of the crystals. In the far field (FF) configuration of QUIP, the transverse momentum anti-correlations shared between the detected and undetected photon are used to realise imaging. The sample and the camera are matched to the Fourier plane of the crystal, with the simplest realisation using a single lens in each arm or parabolic mirror in a 2$f$ configuration. The resulting quantitative imaging performance of QUIP has been comprehensively investigated, both theoretically and experimentally \cite{Lahiri2017, HochrainerPNAS,kviatkovsky2020microscopy,Fuenzalida2022,Vega:2022} in this configuration.

The field of view (FoV) of the undetected light illuminating the sample is determined by the emission angle, $\theta_u$, of the SPDC light. Utilising a single lens of focal length $f$ to match the Fourier plane to the sample plane, we obtain a FoV
 \begin{align}
 {\rm FoV} = 2 f \tan{\theta _u} \approx 2 f \theta _u .
 \end{align}
 The emission angle, $\theta_u$ scales inversely with the square of crystal length as $\theta _u \propto \tfrac{1}{\sqrt{L}}$. Analogous to phase-matching bandwidth, shorter crystals permit more phase-matching angles and accordingly a larger FoV. The resolution in FF QUIP, studied both experimentally and theoretically in detail in \cite{Fuenzalida2022,Vega:2022}, is determined by the strength of momentum anti-correlation shared by the photon pair (detected and undetected) which is ultimately limited by the size of the pump beam illuminating the crystal \cite{Moreau:2018cc}
 \begin{align}
 \delta x = \frac{\sqrt{2 \log{2}} f \lambda _u}{\pi w_p}.
 \end{align}
Other numerical apertures in implementations are rarely a limiting factor, reflecting the fact that crystals are typically much smaller than other optical elements \cite{Fuenzalida2022}. With stronger focusing, the pump photons carry more transverse momenta, which acts to de-correlate the momenta of the emitted photon pair. The number of `spatial modes' available for imaging scales as 
\begin{align}
N \propto \frac{\omega_p}{\sqrt{L(\lambda _u + \lambda _d)}}.
\end{align}
Typical FF configurations using a single lens attain resolutions of between $10-10^3\, \mu$m, where the resolution accessible will scale directly with the illumination wavelength. Most proof-of-concept realisations have thus far considered the near-IR~\cite{lemos2014quantum} and significantly, the mid-IR~\cite{Paterova:2020,kviatkovsky2020microscopy}, where hyperspectral imaging techniques are well motivated. It is important to note that, within the paraxial approximation, the resolution (before magnification) is fixed by the size of the pump waist. When using quasi-phase-matched crystals which are often favoured for their large brightness and collinear phase-matching, the pump waist is constrained by crystal aperture. Further limitations to the imaging performance can arise due to limiting (optical) numerical apertures and aberrations, imperfect alignment, spatial inhomogeneity of the pump, imperfect mode-matching of the beams, or a mismatch in the chosen imaging planes between the crystal, the sample, and the camera \cite{BarretoLemos:22:tutorial}. 
Accessing the resolutions typical of microscopy demands the use of additional optics for magnification, which ideally maintains the available spatial modes while scaling down the size of the illumination spot. This has been typically realised via a three-lens system~\cite{Paterova:2020,kviatkovsky2020microscopy}. This is not a trivial task, as the lens configuration needs to simultaneously satisfy the imaging conditions for both the detected and undetected fields, while also matching the interferometer arm-lengths within the bi-photon coherence length \cite{BarretoLemos:22:tutorial}. Utilising this approach, Petrova {\it et al.} realised a resolution down to 17 $\mu$m at an undetected wavelength of 3$\mu$m, using their system for mid-infrared hyperspectral microscopy of a photosensitive polymer~\cite{Paterova:2020} (Fig.~\ref{fig:microscopyimages}). In a contemporaneous work, Kviatkovsky {\it et al.} imaged a cross-section of a mouse heart at a resolution of 35 $\mu m$ with an illumination wavelength of 3.8 $\mu m$, demonstrating the suitability of the technique for biological imaging. 

The complex morphology of biological samples can cause wavefront distortions, that, owing to the interferometric nature of the technique, mean that the source of the contrast, whether absorption or a phase shift, cannot be distinguished in a single-shot. Kviatkovsky {\it et al.} scanned the interferometer end mirror within the bi-photon coherence length, allowing a pixel-wise reconstruction of the visibility, and consequently the absorption and phase profile of a biological sample~\cite{kviatkovsky2020microscopy}. This approach, akin to quantitative phase microscopy, was also subsequently described in detail in \cite{Fuenzalida2022} and considered in \cite{Topfer2022} where it was termed holography. A different approach that eliminates the need for scanning uses the Wang-Zhou-Mandel interferometer and polarisation optics in detection to implement a phase-quadrature style detection~\cite{Haase:23}.

FF configurations have likely dominated the existing literature because of their experimental ease, relaxed imaging conditions in 2$f$ configurations and clean `illumination'. However, to reach the typical resolutions demanded by microscopy tasks the FF has clear disadvantages. Alongside the optical complexity required to achieve large magnifications while preserving indistinguishability, the Fourier plane suffers from an additional degradation in resolution if the illuminating light is not sufficiently monochromatic~\cite{Devaux2012} -- a clear obstacle for hyperspectral imaging implementations. 

\subsubsection*{Image Plane Imaging}

An alternative pathway to access the small resolutions required of microscopy is to image in the complementary image plane. While it is considered in a comparatively small fraction of the literature, it has been investigated theoretically \cite{Viswanathan:21:1,Viswanathan:21:2} and experimentally \cite{kviatkovsky2022mid,basset2023experimental}. In the image plane, the dependencies previously outlined for far-field imaging essentially invert; we are now concerned instead with the tight spatial correlations that localise where the photons were born in the crystal. In this configuration, the sample and camera are placed in the image plane of the crystal, typically using a 4$f$ system, where the ratio of the lens pair also determines the magnification \cite{kviatkovsky2022mid}. In this scenario, the FoV is directly given by the waist of the pump beam that illuminates the crystal (with an additional lens scaling), defining an aperture within which the SPDC process can occur. 
\begin{align}
FoV = \frac{\sqrt{2\log{2}} w_p}{M}
\end{align}
where $M$ is the magnification of the optical system used to image the undetected light onto the sample. The tightness of the spatial correlations in the so-called photon birth zone scale inversely with the emission angle \cite{Fuenzalida2022}, 
\begin{equation}\label{res}
\delta x =0.41\frac{\lambda_i}{\theta_{i}M}. 
\end{equation}
with the increasing length of the crystal essentially blurring where the photon pair was produced. While the number of available spatial modes remains comparable between the FF and IP, the size of the FoV is typically considerably smaller. With an appropriate choice of the two lenses forming the 4$f$ system, one can access resolutions below 10$\mu$m with comparative experimental ease. Kviatkovsky {\it et al.} demonstrated an image plane re-configuration of their previous experiment \cite{kviatkovsky2022mid}, realising a resolution of 9$\mu$m at an undetected wavelength of 3.8$\mu$m without any evident degradation in the number of spatial modes (see Fig.~\ref{fig:microscopyimages}) .

\subsubsection*{Practical considerations}
In its most fundamental formulation, the imaging performance of a given QUIP system comes down to three parameters: the length of the non-linear crystal, size of the pump waist, and for widely non-degenerate systems, the wavelength of the lowest energy photon (whether detected or undetected)~\cite{Fuenzalida2022}. Crucially, for very long wavelengths, corresponding to the THz-regime ~\cite{Kutas:21}, the number of modes rapidly decreases~\cite{kviatkovsky2022mid}, making point-scanning the only practical imaging modality here.

Taken all together, the central challenge in applicability is the practical trade-off that ultimately arises with wide-field operation: the desire for more spatial modes (and also spectral bandwidth) demands shorter crystals and larger pump waists, which scale poorly in the effective brightness of each spatial mode, requiring longer acquisition times. More pump power and potentially its resonant enhancement~\cite{Lindner2022} offer improvements. Alternatively, accessing the multi-photon regime  via either seeding~\cite{Cardoso2018} or the high peak-powers accessible with pulsed pump light~\cite{hashimoto2023broadband}, offer a path forward.

\subsection{Spectral and 3D imaging}

An alternative approach to imaging is to move away from the widefield operation, where the resolution is ultimately limited by the strength of spatial correlations between signal and idler, and instead utilise a single mode `scanning’ approach. In this approach, the resolution is theoretically limited by the usual optical constraints of classical scanning microscopy. In contrast to widefield imaging, the non-linear interferometer can be operated in a spatially single mode regime, where the pump is comparatively strongly focused and signal and idler emission has a large overlap with the Gaussian fundamental mode. The undetected field is scanned across the sample (or vice-versa), obtaining the position-dependent contrast (absorption/phase) information in a pixel-wise manner. A confocal scanning approach in the near-IR was demonstrated in \cite{Buzas2020} in a SU(2) interferometer seeded at the detection wavelength to improve signal-to-noise. B\'uz\'as {\it et al.} achieved a resolution of 2$\mu$m at an illumination wavelength of 1345 nm, and demonstrated absorption and phase microscopy of different biological samples, including Spirulina filaments and a fruit fly wing~\cite{Buzas2020}. 

This scanning approach is inherently compatible with single-pixel detection and as such, does not necessarily benefit from one of the principal advantages afforded by QUIP - the use of multi-pixel detection technology. However, this approach is compatible with extensions to other sensing modalities, notably optical coherence tomography (OCT) \cite{Valles:2018df} and spectroscopy \cite{Kalashnikov:2016cl}, but also polarimetry~\cite{Paterova:19:Pol} and terahertz sensing \cite{Kutas:21}. 

\subsubsection*{Spectroscopy \& Hyperspectral Imaging} 
The motivation for the widefield imaging results discussed earlier lies in their potential application to hyperspectral imaging tasks, especially in the mid- and far-IR. However, an extension of widefield microscopy to its hyperspectral variant is not straightforward, as the spatial entanglement that facilitates the imaging, and the spectral entanglement that encodes the wavelength information do not, in general, factorise. As a result, the proof-of-principle demonstrations in the near and mid-IR require a relatively narrow detection bandwidth, with multiple detection bands realised either via multiple poling-periods, temperature-tuning \cite{Paterova:2020} or a tuneable filter \cite{kviatkovsky2020microscopy}. One approach would be the wide-field imaging extension of the Fourier transform IR (FTIR) spectroscopy realised in \cite{Lindner:22:1}, where an FTIR spectrum could be extracted for each individual pixel via scanning the interferometer path length~\cite{Placke:23}. 
An alternative approach could utilise a single-mode scanning realisation \cite{Buzas2020}, but implement frequency-resolved detection to realise spectroscopy~\cite{Kalashnikov:2016cl,Kaufmann:22}. The availability of fast and compact CCD grating spectrometers in the visible and near-IR allows for multi-pixel detection, with the typical spectral resolutions down to a few cm$^{-1}$, which is adequate for resolving infrared absorption features in the solid phase \cite{Kaufmann:22} and on par with typical, commercial FTIR spectrometers. 

\subsubsection*{Optical Coherence Tomography} 
Optical coherence tomography (OCT) fills an imaging niche, combining axial imaging high-precision with good spatial resolution. It was one of the first sensing modalities applied to undetected photons~\cite{Valles:2018df,Paterova_2018} and is well-motivated in the infrared, where scattering dramatically decreases with increasing wavelength~\cite{Vanselow:20}. The strong spectral entanglement between the detected and undetected photon pair facilitates the reconstruction of the depth image, with the spectral bandwidth of the bi-photon determining the axial resolution. The first demonstrations of OCT with undetected photons utilised a time-domain (TD) implementation~\cite{Valles:2018df,Paterova_2018} in the near-IR. Vanselow {\it et al.} demonstrated a frequency domain (FD) analog in the mid-IR \cite{Vanselow:20}, utilising a broadband SPDC source and a grating spectrometer, achieving an axial resolution of 10$\mu$m alongside a spatial resolution of 20 $\mu$m at an undetected photon wavelength between 3.3-4.3 $\mu$m. These results were comparable to the current state-of-the-art for classical mid-IR OCT, only limited by the SNR. To overcome the challenges of the weak signals typical with OCT, Machado {\it et al.} implemented OCT with undetected photons in the high-gain regime, dramatically improving the sensitivity~\cite{Machado:2020}. Recent work in this `bright' regime has achieved an axial resolution of 11$\mu$m with broadband illumination between 1450-1650 nm~\cite{hashimoto2023broadband}.

\subsection{Future directions}
\subsubsection*{Multi-photon regime}

One of the primary limitations on the applicability of QIUP and QMIP is the limited brightness available when operating in what we term the `low gain' regime -- where only one photon pair is produced in either of the two crystals. This regime will typically limit sample and/or camera illumination to powers of between a few to 100s of picowatts. In doing so, the timescale of image acquisition is constrained by the shot noise and can limit its applicability in practical applications. An immediate increase in the brightness is available by shifting away from this single biphoton regime, to a multi-photon regime. There are two distinct approaches. The first uses a bright field -- typically a laser -- to seed the non-linear process, which allows for a dramatic improvement in detected brightness~\cite{Shapiro:2015vj,Cardoso2018}. However, seeding an intentionally spatially multi-mode process without degrading the spatial correlations remains a challenge. The second approach instead dramatically increases the gain of the non-linear processes, typically through use of a pulsed, high peak-power pump field. In this regime, the non-linear processes `self-seed', improving in the brightness and ideally the SNR~\cite{WISEMAN2000245,Kolobov_2017}. This regime has already been studied extensively in the context of phase-sensing in non-linear inteferometers, where large gain can overcome the effects of loss. In the context of imaging with undetected light, it has been applied to spectroscopy and OCT ~\cite{Machado:2020,hashimoto2023broadband}. However, a realisation of widefield imaging in the high-gain regime is not as straightforward, as the gain of the non-linear process leads to mixing of the spatial modes and accordingly an effective degradation in mode number.

\subsubsection*{Quantum-enhanced sensing and extreme imaging applications}
Unlike other sensing tasks emerging from quantum ideas, QIUP and QMUP offer, in themselves, no real quantum `advantage'. The advantage offered lies primarily in the ability to decouple the sensing and detection wavelengths with the measurement sensitivities ultimately still bound by the usual shot-noise and diffraction limits~\cite{Vega:2022}. However, there exist recent works~\cite{Kolobov_2017,Miller2021versatilesuper,Black:23} that take inspiration from the field of quantum phase sensing to examine the potential for quantum enhancement. Moreover, Santos {\it et al.} recently proposed a scheme utilising a near-field implementation of QMUP to surpass the diffraction limit~\cite{Santos2022}.

\subsection{Conclusion}
While the fields of QMUP and QIUP are still largely in their infancy, they hold real promise for future applications, whether in their direct application or in inspiring the next generation of technologies. At their best, these technologies offer a simple and cost-effective pathway to sensing at wavelengths where imaging technologies are either non-existent or inferior. This is significant, the development of imaging technologies in the mid- and far-IR has long been constrained by the absence of detection options, despite the unique promise these spectral ranges hold for diagnostics. QMUP presents many desirable characteristics; it circumvents the need for IR light sources and detectors while relying on primarily cost-effective components~\cite{pearce2023practical}. 

While there remains considerable work to transition these approaches from the laboratory to the real world, comparatively simple proof-of-principle demonstrations have spanned illumination wavelengths from the visible to the THz, and achieved resolutions below 10 $\mu$m. Remarkably, in the case of mid-IR OCT, these demonstrations are already competitive with the existing state-of-the-art. In the short term, there is no apparent bottleneck to substantial improvements to the accessible resolutions or number of spatial modes given appropriate optical engineering. There are also several promising pathways to ensure its integration with measurements that can resolve the invaluable spectral information. However, the necessity to realise widefield imaging with an increasing number of spatial modes and improved resolution will demand substantial improvements in the overall brightness.

%
%
%
%

\section{Quantum super-resolution microscopy}

The derivation of the diffraction limit, first presented by Ernst Abbe in his seminal 1873 paper entitled "Contributions to the theory of the microscope and microscopic perception" \cite{abbe1873contributions} served as the basis for the science of quantitative optical microscopy for over a century. With the advent of near-field optics it became clear that the Abbe limit can be broken by violating some of the underlying assumptions (such as that imaging is performed in the far-field \cite{harootunian1986super,durig1986near}). The following decades saw the development of a plethora of methods which utilize 'loopholes' in the Abbe formulation to achieve sub-diffraction limited imaging using far field optics. These methods include stimulated emission depletion microscopy (STED), structured illumination microscopy (SIM), stochastic optical reconstruction microscopy (STORM) and its variants, saturated excitation microscopy (SAX), stochastic optical fluctuation imaging (SOFI) and image scanning microscopy (ISM). Each of these methods is associated with violation of some of the underlying assumptions of the Abbe model. For example, STED and SAX, where either excitation or depletion are saturated, violate the assumption of linearity. Similarly, SIM and ISM utilize multiple images obtained with spatially nonuniform illumination. In STORM and SOFI the imaged scene is not temporally invariant. All of these assumptions are either implicitly or explicitly discussed in the derivation of the diffraction limit.

One implicit assumption clearly made by Abbe is that of classicality. Obviously, the derivation of the diffraction limit predated the dawn of quantum mechanics by several decades. As such, the combination of imaging and quantum optics is, at least at the conceptual level, a natural one in the pursuit of superresolution. Fortunately, in the past decade or so, the experimental tools available have been able to support several demonstrations of quantum implementations towards superresolution, and this arena is rapidly growing.

The term 'quantum super-resolution microscopy' is in some sense misleading as it relates to a broad range of ideas, each associated with a different aspect of quantumness or with different quantum phenomena. As such, it is convenient to divide the discussion below to three separate topics: (1) Quantum resolution limits in model-based superresolution, (2) Microscopy enhanced by quantum photon statistics and (3) Microscopy enhanced by quantum entanglement of photon pairs. Each of these topics represents a different approach to the utilization of quantum properties of light towards achieving superresolution imaging under a certain set of assumptions. These approaches were first presented as theoretical frameworks, but the dramatic advances in single-photon detector technologies over the past decade now render them experimentally accessible. In the following, we first provide an overview of each of these research directions and then wrap up with a short discussion about the technology required to promote them towards practical imaging applications. 

\subsection{Model-based quantum super-resolution}
In its simplest realization, model-based quantum super-resolution is not an imaging modality but rather the identification of an optimal solution of an estimation problem. It can find its roots in the field of compressed sensing, whereby sparse data sets are adequately reconstructed without fulfilling the Nyquist criterion. The problem of providing a good estimation of an arbitrarily small distance between two independent emitters has been a cornerstone of this branch of quantum-inspired super-resolution methods. Deconvolution can be used to differentiate between a single emitter and a pair of spatially separated emitters (when the point spread function is known and the signal-to-noise ratio is sufficient).
However, deconvolution becomes notoriously difficult as the lateral distance between emitters becomes smaller. A general framework to approach this problem, utilizing the quantum Cramer-Rao bound, was proposed to show that upon appropriate measurements estimating emitter separation turns into a solvable problem at any distance between the two emitters \cite{tsang2016quantum}. Essentially, this means that the real-space basis is far from being an optimal one to estimate the distance between two separate uncorrelated emitters. This was rapidly followed by several experimental demonstrations of this idea, which vary in the detailed implementation (mode sorting, interferometry with an inverted image etc.) but follow similar logic \cite{paur2016j,tang2016fault,yang2016far,tham2017beating}. Further extensions of this idea to other scenarios were later demonstrated, for example estimation of the distance between two axially separated emitters \cite{zhou2019quantum}. A series of recent works have tried to alleviate the restrictions on the optimization problem so as to accommodate a broader range of emitter distributions while taking advantage of the different measurement basis set. These have shown improvements over conventional deconvolution in both confocal \cite{Bearne2021} and widefield imaging \cite{Frank2023} scenarios.

Notably, none of the above actually utilizes quantum resources. Rather, ideas are borrowed from the quantum world and implemented classically via analogy with the estimation of quantum wavefunctions. Yet, while the problem as posed may seem artificial (i.e. estimation of the distance between two isolated emitters), these ideas can potentially be adopted in new realizations of superresolution microscopy schemes, where quantum information is used in combination with localization based microscopy. Localization super-resolution microscopy typically operates in the regime of substantially less than one active emitter per diffraction limited spot, so as to avoid false localizations when two or more emitters are present simultaneously. This places a significant burden on the total measurement time for superresolved imaging. While better localization algorithms can overcome, to some extent, situations where emitter density is increased, they likely fail at very high emitter densities. An experimental estimate of the number of emitters within a diffraction limited spot can assist localization at higher densities. This can be reliably performed only using quantum photon statistics.

Emitter number estimation utilizes the fact that most fluorophores used in bioimaging are 'quantum emitters', that is they can only emit one photon at a time \cite{fleury2000nonclassical}. This phenomenon is typically characterized using a Hanbury Brown and Twiss detection setup, splitting the emitted photon stream between two (or more) detectors and looking at the correlations within the photon stream. Single emitters are characterized by `photon antibunching', the absence (or at least scarcity) of simultaneous photon detection events. This is reflected as a dip in the second-order temporal autocorrelation function $g^{(2)}(\tau)$ which reaches zero at $\tau=0$. For ensembles of N identical independent emitters this value increases to $g^{(2)}(0) = 1-\frac{1}{N}$, a signature which has been used in the development of the counting by photon statistics (CoPS) method \cite{weston2002measuring,ta2010experimental,grussmayer2017time} to estimate the number of up to 30 emitters. Notably, for large numbers of emitters higher order moments (e.g. the third order autocorrelation function $g^{(3)}(\tau_1,\tau_2)$ describing correlations between three detection channels) can be used.

Emitter number estimates alleviate the need to work at low active emitter densities in STORM-like scenarios, as diffraction limited spots with multiple active emitters can be detected and rejected. The higher active emitter densities this approach enables can substantially enhance localization microscopy's acquisition speeds. Alternatively, one can imagine performing localization microscopy also on spots with multiple emitters, utilizing the known emitter number and the available tools for emitter separation estimation (the mean position can, in principle, be detected via centroid estimation). There have been reports of some very crude implementations of such quantum-assisted localization imaging. Estimation of the distance between exactly two emitters (as determined from photon statistics) down to circa 10nm was demonstrated using NV centers in diamond \cite{cui2013quantum}. Estimation of the number of emitters in a diffraction limited spot prior to STED imaging (which can, in principle, facilitate imaging by choosing the appropriate resolution) was also demonstrated \cite{ta2015mapping}. Emitter counting and localization using a combination of classical and quantum fluctuations was used to avoid false localizations in STORM imaging using a fiber bundle as a fast camera (with a very small field of view) \cite{israel2017quantum}. Yet, these are still far from realizing the full potential imbued in knowledge of photon statistics, likely because this requires tools for rapid, wide-field characterization of photon statistics which are only now becoming more available. These include large format (circa one megapixel) single photon detector arrays and particularly silicon-based SPAD arrays.

\subsection{Super-resolution via direct use of quantum photon statistics}
Photon antibunching can be directly used as a resource for enhancing the optical resolution of an imaging system. In fact, antibunching is a quantum analog of stochastic classical fluctuations which have previously been utilized for enhancing the resolution of optical microscopes via methods such as SOFI \cite{dertinger2009fast}. Both classical and quantum fluctuations correspond to deviations from Poisson statistics of the number of photons emitted per unit time. Classical fluctuations, such as fluorophore blinking, lead to super-Poissonian statistics whereas antibunching leads to sub-Poissonian statistics. Mathematically, however, it is the deviation from Poisson statistics that matters, as it introduces higher moments of the distribution that are not determined by the mean photon count. The deviation of the Nth moment of the distribution from Poisson statistics, as characterized by the correlation between detections from N detectors, is associated with a point spread function (PSF) which corresponds to the intensity PSF raised to the power of N. Naively, this leads to a resolution which is about $\sqrt{N}$ better than the diffraction limit. Yet, the Fourier support is N times broader such that with appropriate Fourier reweighting (and sufficient SNR for the high frequencies) a resolution increase of up to N times beyond the diffraction limit is possible.

The idea of using photon antibunching is closely related to the concept of centroid estimation using multiple photons simultaneously emitted by the same emitter. This idea was first presented in a thought experiment where emitters which always emit photons in simultaneous pairs and infinitely fast detection hardware were assumed \cite{hell1995two}. In this case, pair detection provides two independent estimates of the emitter position and thus resolution is enhanced by a factor of $\sqrt{2}$. Notably, however, multiple-photon emission represents super-Poissonian emission statistics (therefore the light is not 'quantum'). Realization of this experiment proved difficult in the absence of appropriate emitters, although centroid estimation has been demonstrated using SPDC photon pairs in transmission mode imaging \cite{unternahrer2018super,toninelli2019resolution}. In contrast, emitters which exhibit antibunching are abundant, and can be thought of as emitters of 'missing pairs' in the photon stream. The difference is that 'missing pairs' are detected against a bright background and thus suffer from a non-favorable scaling in the presence of noise.

The main hurdle towards the implementation of antibunching-based superresolution imaging has been the availability of suitable detectors. As antibunching estimation requires temporally correlating single-photon detections, many binary frames (a photon detected or not at each pixel at a given time) are required to form a single image. Thus, for wide-field imaging, a single-photon sensitive imaging sensor with a high frame rate (preferably ${>}$ 1MHz) is necessary, a system which is currently unavailable. The first demonstration of this method thus relied on an EMCCD detector operated at Geiger mode at a modest readout rate of 1kHz \cite{schwartz2013superresolution}. For this to work, the sample had to be illuminated by pulsed excitation at a rate of one pulse per frame, with the added requirement that the excitation pulses are much shorter than the radiative decay time of the emitters to avoid multiple excitation-emission cycles within a single excitation pulse. Under these conditions, many minutes were needed to obtain an antibunching-based superresolved image using second-order correlations and an hour was needed to obtain very noisy images with third-order correlations. With the availability of much faster smaller format single photon imaging detectors (a fiber bundle camera or a SPAD array), antibunching was used to augment the spatial resolution in confocal imaging \cite{monticone2014beating} and in image scanning confocal microscopy \cite{tenne2019super,lubin2019quantum} - achieving circa threefold resolution increase beyond the diffraction limit for the latter (by combining the resolution increase from ISM and that from photon statistics). Antibunching-based imaging can in principle be combined with other superresolution modalities using structured illumination such as SIM \cite{classen2017superresolution}, random illumination microscopy \cite{liu2022resolution} or SPIFI \cite{field2016superresolved}. The former two are still impractical in the absence of suitable detectors. The latter, which can use a single-pixel detector, can be implemented with current day technology.

Beyond the technological aspects, several key points regarding the limitations of antibunching based imaging are important. The first relates to the practicality of the use of higher (${>2}$) order correlations to further enhance the imaging resolution. While this is possible (and has already been demonstrated \cite{schwartz2013superresolution}), it is important to realize that the signal-to-noise of higher order antibunching scales unfavorably with the emitter density. While the antibunching signal scales linearly with the number of emitters, the shot noise of the background against which it is measured scales as the number of emitters to the power N/2, where N is the order of the correlation signal. Thus, for densely labeled scenes, antibunching based imaging is limited to second order. The second relates to imaging time. As 'missing pairs' are detected less frequently than fluorescence photons, longer imaging times are needed to obtain a decent SNR. This is mostly of interest in scanning microscopy, where various schemes for algorithmic acceleration of imaging by combining both the classical and the quantum signals have been implemented \cite{rossman2019rapid}. Finally, since antibunching based imaging requires excitation at some level of saturation, it is more suitable for use with fluorescent probes which do not suffer from rapid photobleaching.

\subsection{Super-resolution via the use of entangled photon pairs}
Photon entanglement is a much more potent resource than photon statistics alone, and has enabled a plethora of applications as discussed in previous chapters. Yet, in the context of superresolution microscopy the implementation of entanglement-based schemes is still lagging behind, with only a handful of realizations of such principles. The important point to note is that to exploit the full benefit of entanglement one must go beyond taking advantage only of the simultaneous arrival of multiple photons (which essentially is reflected in the statistics).

The problem of exploiting entanglement to perform superresolution imaging is intimately linked to that of superresolved lithography. The use of photon-number states (N00N states) for superresolved lithography via the mechanism of enhanced N-photon absorption, as discussed above, can enable the writing of sub-Abbe limit features \cite{boto2000quantum}. These can be directly linked to superresolved imaging via centroid estimation (experimentally demonstrated to second order \cite{shin2011quantum} and to higher orders \cite{rozema2014scalable}).

Another avenue for exploiting entanglement towards superresolution has been enabled through the use of holographic methods where phase information is encoded in the second order coherence \cite{defienne2021polarization}. In this scheme, momentum and polarization entangled photons are used to encode a phase image, where the phase is encoded in the polarization state and the position in the momentum. Readout is performed by measuring photon pair spatial correlations between entangled momentum components upon application of a relative phase between the two polarization states with a spatial light modulator. As the second order coherence is used here, the maximal spatial resolution is twice that afforded by the diffraction limit for an equivalent intensity image. A major advantage of using polarization and momentum entangled photons is that it enables wide-field imaging using a standard EMCCD camera, as compared with scanning approaches. A similar approach has also been used to overcome limits due to detector pixellization \cite{defienne2022pixel}. 

\subsection{Conclusion}

Far-field superresolution microscopy has turned from a dream to a reality, having a significant impact on life science research, in less than 30 years. This is in part due to the fact that some of the superresolution modalities are readily implementable on relatively standard microscope platforms that are accessible to the non-expert user. Harnessing quantum optical phenomena on such standard microscope systems would have seemed far-fetched only a few years ago, especially in light of the very modest improvement in spatial resolution offered by quantum superresolution schemes \cite{boto2000quantum,schwartz2013superresolution,defienne2021polarization}. Yet, two things have the potential to change this. The first is the realization that quantum optics can be used in conjunction with 'classical' modalities (e.g. STORM, SIM, ISM) to enhance the performance of these methods without affecting the usability of the method for the user. This performance enhancement is not limited to resolution alone, as it can also support a dramatic reduction of acquisition times for localization microscopy. The second relates to advances in hardware development, and particularly the dramatic advances in single-photon sensitive array detectors which are necessary components in the practical implementation of quantum optics schemes (requiring measurement of photon correlations) in an imaging platform which is compatible with current day microscopes. Large format array detectors are now in advanced stages of development or even available on the market for PMTs, APDs and superconducting nanowire detectors, which should provide access to a broad wavelength range while maintaining an excellent time resolution. Clearly, additional progress in hardware (and especially the shift towards CMOS-compatible systems) is necessary to provide sufficient performance, mostly in terms of fill factor, detection efficiency and crosstalk reduction for these systems to be attractive alternatives to high performance sCMOS and CCD detectors available today. In parallel, better ways to deal with the high data rates obtained in single-photon measurements so as to enable real-time operation and analysis should be developed. Considering these challenges, however, the benefits of improved resolution (in conjunction with additional benefits of photon correlation measurements such as the high time resolution and quantitative fluorescence measurements to the level of emitter number estimation) seem sufficiently lucrative to further pursue this avenue.

\section{Conclusion}

In this review paper we have sought to provide an accessible overview to the field of quantum light microscopy, to what has been achieved, what the opportunities for the future are, what challenges remain and what the practical limitations are. Microscopes are a primary tool applied to our understanding of the microscopy processes that govern all life. Advances in microscopy have time and again led to new discoveries and surprises that have transformed our understanding of life and driven major societal benefits across healthcare, pharmaceuticals, biotechnology, agriculture and many other areas. Quantum-engineered light provides a new tool for microscopy. Its use in quantum light microscopes has enabled new advances and promises to drive new impact. We hope that this review paper stimulates interest in this new form of microscope, and contributes toward the acceleration of this exciting new field.

\section*{Acknowledgements}

W.P.B. and A.T. acknowledge support from the Air Force Office of Scientific Research under award numbers FA9550-20-1-0391 and FA9550-22-1-0047, the Australian Research Council Centre of Excellence for Engineered Quantum Systems (EQUS, CE170100009), and the Australian Research Council Centre of Excellence in Quantum Biotechnology (QUBIC, CE230100021). D.T. and R.T. acknowledge support from the Swiss National Science Foundation through the Sinergia grant CRSII5-170981 and QuantERA E$^2$TPA grant 200020\_213126. D.O. acknowledges support from the Israeli Science Foundation and the Directorate for Defense Research and Development (DDR\&D), grant No. 3415/21, and from the Israel Science Foundation data science program.

\section*{Author biographies}

\subsection*{Warwick P. Bowen}

Prof Bowen’s research focuses on the implications of quantum science on precision measurement, and applications of quantum measurement in areas ranging from quantum condensed matter physics to the biosciences. He is a Fellow of the Australian Institute of Physics and Director of the Australian Centre of Excellence in Quantum Biotechnology. 

\subsection*{Helen M. Chrzanowski}
Helen Chrzanowski is an experimental physicist, specialising in the study of the quantum properties of light, especially as a medium to measure, communicate and process information. After a PhD at the Australian National University in Canberra in continuous variable quantum optics, Helen undertook postdoctoral positions at the University of Oxford and the Humboldt University in Berlin. 

\subsection*{Dan Oron}

Dan Oron earned a B.Sc. in mathematics and physics from the Hebrew university in 1994. He earned his M.Sc. degree in physics from Ben-Gurion University of the Negev in 1998 and received his Ph.D., also in physics, from the Weizmann Institute of Science in 2005, under the guidance of Prof. Yaron Silberberg. After conducting postgraduate research with Prof. Uri Banin at the Hebrew University for two years, he joined the staff of the Weizmann Institute in April 2007. He is currently a professor at the department of Molecular Chemistry and Materials Science at the Weizmann institute. His main research interests are at the interface between light and the nanoscale, studying both the interaction of light with nanostructured materials (mostly inorganic and hybrid semiconductor nanocrystals), optical superresolution methods harnessing both quantum and classical fluctuations in light emission and the optics of biological nanostructured materials.

\subsection*{Sven Ramelow}

Dr. Sven Ramelow is for more than 15 years involved in experimental research with single photons. After his studies of physics at Humboldt-University Berlin, he completed his PhD and first PostDoc in Prof. Anton Zeilinger’s group at the University of Vienna. Here he dedicated his research to a number of fundamental quantum experiments as well as application-motivated topics like quantum communication and quantum imaging. During his second PostDoc at Cornell University, USA in Prof. Alex Gaeta’s group he extended his expertise in integrated quantum optics and quantum frequency conversion. Since 2017 Dr. Ramelow is leading an Junior Research Group at Humboldt-University Berlin experimentally working on integrated optics, quantum frequency conversion and SPDC-based imaging, spectroscopy and OCT with undetected mid-IR photons.

\subsection*{Dmitry Tabakaev}

Dr. Dmitry Tabakaev is a Scientist in the Photonic systems Research unit at Silicon Austria Labs. He did his PhD and Postdoc on the topic of entangled two-photon absorption, single-photon fluorescence and phosphorescence detection. Currently his field of research spans from laser interferometry to surface acoustic waves to quantum nonlinear optics of novel metasurfaces.

\subsection*{Alex Terrasson}

Alex Terrasson is an early-career researcher in the field of quantum optics and biosensing. He earned a Master's Degree in nanotechnology and quantum physics in 2017 from Ecole Centrale Lyon. In 2023, he successfully defended his PhD at the University of Queensland in Quantum Optics under the supervision of Prof. Warwick Bowen. His doctoral research focused on quantum-enhanced Raman imaging and optical tweezers applied to viscosity and biology measurements. In July 2023, Alex assumed the role of a postdoctoral fellow with Prof. Bowen to further his research in quantum biosensing. 

\subsection*{Rob Thew}
Rob Thew is a senior researcher and group leader in the Quantum Technologies group at the University of Geneva. Rob Thew works in the field of quantum communication and sensing, spanning fundamental to applied topics, and from technology development to systems integration. He was the inaugural chair of the Strategic Research Agenda Work Group in the context of the European Quantum Flagship initiative. He is also the founding Editor-in-Chief for the IOP journal: Quantum Science and Technology. 


\bibliographystyle{tfnlm}
\bibliography{interactnlmsample}

\end{document}